# Repurpose Open Data to Discover Therapeutics for COVID-19 using Deep Learning

Xiangxiang Zeng[1,#], Xiang Song[2,#], Tengfei Ma[1], Xiaoqin Pan[1], Yadi Zhou[3], Yuan Hou[3], Zheng Zhang[2,4], George Karypis[5,6,*], Feixiong Cheng[3,7,8,*]

[1]School of Computer Science and Engineering, Hunan University, Changsha, 410012, China

[2]AWS Shanghai AI Lab, Shanghai 200335, China

[3]Genomic Medicine Institute, Lerner Research Institute, Cleveland Clinic, Cleveland, OH 44106, USA

[4]New York University Shanghai, Shanghai 200122, China

[5]AWS AI, East Palo Alto 94303, CA, USA

[6]Department of Computer Science and Engineering, University of Minnesota 200 Union Street SE, Minneapolis, MN, USA

[7]Department of Molecular Medicine, Cleveland Clinic Lerner College of Medicine, Case Western Reserve University, Cleveland, OH 44195, USA

[8]Case Comprehensive Cancer Center, Case Western Reserve University School of Medicine, Cleveland, OH 44106, USA

[#]These authors are joint first authors on this work.

*To whom correspondence should be addressed:

Feixiong Cheng, PhD

Lerner Research Institute

Cleveland Clinic

Email: chengf@ccf.org

George Karypis, PhD

Distinguished McKnight University Professor

University of Minnesota

Email: karypis@umn.edu

## Abstract

There have been more than 850,000 confirmed cases and over 48,000 deaths from the human coronavirus disease 2019 (COVID-19) pandemic, caused by novel severe acute respiratory syndrome coronavirus (SARS-CoV-2), in the United States alone. However, there are currently no proven effective medications against COVID-19. Drug repurposing offers a promising way for the development of prevention and treatment strategies for COVID-19. This study reports an integrative, network-based deep learning methodology to identify repurposable drugs for COVID-19 (termed CoV-KGE). Specifically, we built a comprehensive knowledge graph that includes 15 million edges across 39 types of relationships connecting drugs, diseases, genes, pathways, and expressions, from a large scientific corpus of 24 million PubMed publications. Using Amazon's AWS computing resources, we identified 41 repurposable drugs (including indomethacin, toremifene and niclosamide) whose therapeutic association with COVID-19 were validated by transcriptomic and proteomic data in SARS-CoV-2 infected human cells and data from ongoing clinical trials. While this study, by no means recommends specific drugs, it demonstrates a powerful deep learning methodology to prioritize existing drugs for further investigation, which holds the potential of accelerating therapeutic development for COVID-19.

**Introduction**

As of April 22, 2020, there have been more than 850,000 confirmed cases and over 48,000 deaths from Coronavirus Disease 2019 (COVID-19), the disease caused by the virus SARS-CoV-2, in the United States alone.[1] However, there are currently no proven effective antiviral medications against COVID-19.[2] There is an urgent need for the development of effective treatment strategies for COVID-19. Traditional *de novo* drug discovery was estimated that pharmaceutical companies spent $2.6 billion in 2015 for the development of an FDA-approved new chemical entity drug.[3] Drug repurposing, a drug discovery strategy from existing drugs, offers an promising way for the development of prevention and treatment strategies for COVID-19.[4]

In a randomized, controlled, open-label trial[5], lopinavir and ritonavir combination therapy did not show clinical benefit compared to standard care for hospitalized adult patients with severe COVID-19, limiting the traditional antiviral treatment for COVID-19. SARS-CoV-2 replication and infection depends on the host cellular factors (including angiotensin-converting enzyme 2 (ACE2)) for entry into cells.[6] Systematic identification of virus-host protein-protein interactions (PPIs) offers an effective way towards elucidation of the mechanisms of viral infection; furthermore, targeting cellular virus-host interactome offers a promising strategy for the development of effective drug repurposing for COVID-19 as demonstrated in previous studies.[7-9] We recently demonstrated that network-based methodologies leveraging relationship between drug targets and diseases can serve as a useful tool for efficient screening of potentially new indications for FDA-approved drugs with well-established pharmacokinetics/pharmacodynamics, safety and tolerability profiles.[10-12] Yet, without foreknowledge of the complex networks connecting drugs, targets, SARS-CoV-2 and diseases, development of affordable approaches for effective treatment of COVID-19 is challenging.

Prior knowledge of networks from large scientific corpus of publications offers a deep biological perspective for capturing the relationships between drugs, genes and diseases (including COVID-19). Yet, extracting connections from a large-scale repository of structured medical information is challenging. In this study, we present the state-of-the-art knowledge graph-based, deep learning methodologies for rapid discovery of drug candidates to treat COVID-19 from 24 million PubMed publications. Via systematic validation using transcriptomics and proteomics data generated from SARS-CoV-2 infected human cells and the ongoing clinical trial data, we successfully identified 41 drug candidates that can be further tested in large-scale randomized control trials for potential treatment of COVID-19.

## Results

**Pipeline of CoV-KGE**

Here we present a knowledge graph-based, deep learning methodology for drug repurposing in COVID-19, termed CoV-KGE (**Figure 1**). Our method uses DGL-KE, developed by our Amazon's AWS AI Laboratory[13], to efficiently learn embeddings of large knowledge graphs. Specifically, we construct a knowledge graph from 24 million PubMed publications[14] and DrugBank[15], including 15 million edges across 39 types of relationships connecting drugs, diseases, genes, anatomies, pharmacologic classes, gene/protein expression, and others (*cf.* **Supplementary Tables 1 and 2**). In this knowledge graph, we represent the Coronaviruses (CoVs) by assembling multiple types of known CoVs, including SARS-CoV-1, MERS-CoV, HCoV-229E, and HCoV-NL63 as described in our recent study[9].

We next utilize DGL-KE's knowledge graph embedding model, RotatE[16], to learn representations of the entities (e.g., drugs and targets) and relationships (e.g., inhibition relation between drugs and targets) in an informative, low-dimensional vector space. In this space, each relationship type (e.g., antagonists or agonists) is defined as a rotation from the source entity (e.g., hydroxychloroquine) to the target entity (e.g., toll-like receptor 7/9 (TLR7/9)) (*cf.* Methods).

**High performance of CoV-KGE**

After mapping drugs, CoVs, and the *treatment* relationships to a complex vector space using RotatE, the top-100 most relevant drugs are selected as candidate for CoVs in the *treatment* relation space **(Supplementary Figure 1)**. Using the ongoing COVID-19 trial data (https://covid19-trials.com/) as a validation set, CoV-KGE has the larger area under the receive operating characteristic curve (AUROC = 0.85, **Figure 2**) in identifying repurposable drugs for COVID-19.

We next employ t-SNE (t-distributed stochastic neighbor embedding algorithm[17]) to further investigate the low-dimensional node representation learned by CoV-KGE. Specifically, we project drugs grouped by the first-level of the Anatomical Therapeutic Chemical (ATC) classification systems code onto two-dimensional space. **Figure 3A** indicates that CoV-KGE is able to distinguish 14 types of drugs grouped by ATC codes, consistent with a high AUROC value of 0.85 (**Figure 2**).

We further validate the top candidate drugs using enrichment analysis of drug-gene signatures and SARS-CoV induced transcriptomics and proteomics data in human cell lines (*cf.* Methods). Specifically, we analyzed three transcriptomic datasets in SARS-CoV-1 infected human cell lines and one proteomic dataset in SARS-CoV-2 infected human cell lines. In total, we obtained 41 high-confidence repurposable drugs (**Supplementary Table 3**) using subject matter expertise based on a combination of factors: (i) strength of the CoV-KGE predicted score; (ii) availability of clinical evidence from ongoing COVID-19 trials; (iii) availability and strength of enrichment analyses from SARS-CoV-1/2 effected human cell lines. Among the 41 candidate drugs, 11 drugs are or have been under clinical trials for COVID-19, including thalidomide, methylprednisolone, ribavirin, umifenovir, chloroquine, hydroxychloroquine, tetrandrine, suramin, dexamethasone, lopinavir, azithromycin (**Figure 3A** and **Supplementary Table 3**). We next turn to highlight three types of predicted drugs for COVID-19, including anti-inflammatory agents, selective estrogen receptor modulators (SERMs), and antiparasitics (**Figure 3A**).

**Discovery of drug candidates for COVID-19 using CoV-KGE**

Given the well-described lung pathophysiological characteristics and immune responses (cytokine storms) of severe COVID-19 patients, drugs that dampen the immune responses may offer effective treatment approaches for COVID-19.[18, 19] As shown in **Figure 3A**, we computationally identified multiple anti-inflammatory agents for COVID-19, including indomethacin and melatonin. Indomethacin, an approved cyclooxygenases (COXs) inhibitor, has been widely used for its potent anti-inflammatory and analgesic properties[20]. Indomethacin has been reported to have antiviral properties, including SARS-CoV-1[20], and SARS-CoV-2[21]. Importantly, preliminary *in vivo* observation showed that oral indomethacin (1 mg/kg body weight daily) reduced the recovery time of SARS-CoV-2 infected dogs.[21] Melatonin plays a key role in the regulation of human circadian rhythm that alters the translation of thousands of genes, including melatonin-mediated anti-inflammatory and immune-related effects for COVID-19. Melatonin has various antiviral activities by suppressing multiple inflammatory pathways[22, 23] (including IL6, IL-1β, and TNFα, etc.) -- these inflammatory effects are directly relevant given the well-described lung pathophysiological characteristics of severe COVID-19 patients. The melatonin's mechanism-of-action may also help explain the epidemiologic observation that children, who have naturally high melatonin levels, are relatively resistant to COVID-19 disease manifestations, while older individuals, who have decreasing melatonin levels with age, are a very high-risk population.[24] In addition, exogenous melatonin administration may be of particular benefit to older patients given an aging-related reduction of endogenous melatonin levels and vulnerability of older individuals to the lethality of SARS-CoV-2.[24]

An overexpression of estrogen receptor has played a crucial role in inhibiting viral replication and infection.[25] Several selective estrogen receptor modulators (SERMs), including clomifene, bazedoxifene, and toremifene, are identified as promising candidate drugs for COVID-19 (**Figure 3B**). Toremifene, the first generation of nonsteroidal SERM, was reported to block various viral infections at low micromolar concentration, including Ebola virus[26, 27], MRES-CoV[28], SARS-CoV-1[29], and SARS-CoV-2[30] (**Figure 3A**). Toremifene prevents fusion between the viral and endosomal membrane by interacting with and destabilizing the virus glycoprotein, and eventually blocking replications of the Ebola virus.[26] The underlying antiviral mechanisms of SARS-CoV-1 and SARS-CoV-2 for toremifene remains unclear and are currently being investigated. Toremifene has been approved for the treatment of advanced breast cancer[31] and has also been studied in men with prostate cancer (~1,500 subjects) with reasonable tolerability.[32] Toremifene is 99% bound to plasma protein with good bioavailability and typically administered at a dosage of 60 mg orally.[33] In summary, toremifene offers a promising candidate drug with ideal pharmacokinetics properties to be tested in COVID-19 clinical trials directly.

Hydroxychloroquine and chloroquine phosphate, two approved antimalarial drugs, though lacking strong clinical evidence, were authorized by the U.S. FDA for treatment of COVID-19 patients using emergency use authorizations.[2] In this study, we identified that both hydroxychloroquine and chloroquine are among the predicted candidates for COVID-19 (**Figure 3A**). Between the two, hydroxychloroquine's *in vitro* antiviral activity against SARS-CoV-2 is stronger than that of chloroquine (hydroxychloroquine: 50% effective concentration ($EC_{50}$) = 6.14 µM where chloroquine: $EC_{50}$ = 23.90 µM)[34]. Hydroxychloroquine and chloroquine are known to increase the pH of endosomes, which inhibits membrane fusion, a required mechanism for viral entry (including SARS-CoV-2) into the cell[19]. Additionally, the potential benefit of hydroxychloroquine in COVID-19 patients may be explained by both antiviral and anti-inflammatory effects (**Figure 3B**). *In vitro* experiments showed that hydroxychloroquine reduced the expression of vascular cell adhesion molecule 1 (VCAM1) and IL-1β and attenuates the suppression of endothelial nitric oxide synthase (NOS3) in human aortic endothelial cells.[10] However, even though chloroquine and hydroxychloroquine are relatively well tolerated, several adverse effects (including QT prolongation) limit its clinical use for COVID-19 patients, especially for patients with pre-existing cardiovascular disease or diabetes.[10, 35-37] These findings suggest that hydroxychloroquine have both anti-inflammatory and antiviral activity for COVID-19 therapy but its safety and efficacy needs to be tested in large-scale randomized control trials. In addition, niclosamide, an FDA-approved drug for treatment of tapeworm infestation, was recently identified to have a stronger inhibitory activity on SARS-CoV-2 at sub-micromolar ($IC_{50}$ = 0.28 μM). Gassen et al. showed that niclosamide inhibited SKP2 activity by enhancing autophagy and reducing MERS-CoV replication as well.[38] Altogether, niclosamide offers another promising drug candidate for COVID-19, which is warranted to be investigated in randomized controlled trials further.

Finally, given the up-regulation of systemic inflammation—in some cases culminating to a cytokine storm observed in severe COVID-19 patients[40]—combination therapy with an agent targeting inflammation (melatonin or indomethacin) and with direct antiviral effects (toremifene and niclosamide) has the potential of leading to successful treatments.

**Discussion**

As COVID-19 patients flood hospitals worldwide, physicians are trying to search effective antiviral therapies to save lives. Multiple COVID-19 vaccine trials are under way; yet, it might not be physically possible to make enough vaccine for everyone in a short period of time. Furthermore, SARS-CoV-2 replicates poorly in multiple animals, including dogs, pigs, chickens, and ducks, which limits preclinical animal studies[39].

In this study, we present CoV-KGE, a powerful, integrated AI methodology that can be used to quickly identify drugs that can be repurposed for the potential treatment of COVID-19. Our approach can minimize the translational gap between preclinical testing results and clinical outcomes, which is a significant problem in the rapid development of efficient treatment strategies for the COVID-19 pandemic. From a translational perspective, if broadly applied, the network tools developed here could help develop effective treatment strategies for other emerging infectious diseases and other emerging complex diseases as well.

## Methods and Materials

**Constructing the knowledge graph**

In this study, we constructed a comprehensive knowledge graph (KG) from GNBR[14] and DrugBank[41]. First, from GNBR, we included in the KG relations corresponding to drug-gene interactions, gene-gene interactions, drug-disease associations, and gene-disease associations. Second, from DrugBank, we selected the drugs whose molecular mass is greater than 230 daltons and also exist in GNBR—resulting in 3,481 FDA-approved and clinically investigational drugs. For these drugs, we included in the KG relationships corresponding to the drug-drug interactions, and the drug-side-effects, drug ATC codes, drug-mechanism-of-action, drug-pharmacodynamics, and drug-toxicity associations. Third, we included the experimentally discovered coronaviruses-gene relationships from our recent work in the KG.[9] Fourth, we treated the COVID-19 context by assembling six types of Coronaviruses (including SARS-CoV, MERS-CoV, HCoV-229E, and HCoV-NL63) as a comprehensive node of Coronaviruses (CoVs), and rewired the connections (edges) from genes and drugs. The resulting KG contains four types of entities (drug, gene, disease, and drug side information), 39 types of relationships (Supplementary **Table 1),** 145,179 nodes, and 15,018,067 edges (Supplementary **Table 2)**.

**The knowledge graph embedding model RotatE**

Models for computing knowledge graph embeddings learn vectors for each of the entities and each of the relation types so that they satisfy certain properties. In our work we learned these vectors using the RotatE model.[42] Given an edge in the knowledge graph represented by the triplet (*head-entity, relation-type, and tail-entity*), RotatE defines each relation type as a rotation from the head entity to tail entity in the complex vector space. Specifically, if $h, t$ are the vectors corresponding to the head and tail entities, respectively and $r$ is the vector corresponding to the relation type, RotaE tries to minimize the distance

$$d_r(h, t) = ||h \circ r - t|| \qquad (1)$$

where $\circ$ denotes the Hadamard (element-wise) product.

In order to minimize the distance between the head and the tail entities of the existing triplets (positive examples) and maximize the distance among the non-existing triples (negative examples), we use the loss function

$$L = -\log \sigma\big(\gamma - d_r(h, t)\big) - \sum_{i=1}^{n} p(h_i, r, t_i) \log \sigma(d_r(h_i, t_i) - \gamma) \qquad (2)$$

where $\sigma$ is sigmoid function, $\gamma$ is a margin hyperparameter with $\gamma > 0$, $(h_i, r, t_i)$ is a negative triplet, and $p(h_i, r, t_i)$ is the probability of occurrence of the corresponding negative sample.

**Details of DGL-KE package**

DGL-KE[43] is a high performance, easy-to-use, and scalable package for learning large-scale knowledge graph embeddings with a set of popular models including TransE, DistMult, ComplEx, and RotatE. It includes various optimizations that accelerate training on knowledge graphs with millions of nodes and billions of edges using multi-processing, multi-GPU, and distributed parallelism. DGL-KE is able to compute the RotatE-based embeddings of our knowledge graph in around 40 minutes on an EC2 instance with 8 GPUs under Amazon's AWS computing resources.

**Experimental Settings**

We divide the triplets (e.g., a relationship among drug, treatment, and disease) into training set, validation set, and test set in a 7:1:2 manner. We selected the embedding dimensionality of dim=200 for nodes and relations. The RotatE is trained for 16000 epochs with batch size 1024 and 0.1 as the learning rate. We choose $\gamma = 12$ as the margin of the optimization function in Supplementary Note 2.

**Gene-Set Enrichment Analysis**

We performed the gene set enrichment analysis to further validate the predicted drug candidates from CoV-KGE. We first collected three differential gene expression data sets of human cell lines infected by HCoVs from the Gene Expression Omnibus database (https://www.ncbi.nlm.nih.gov/geo/). Specifically, two transcriptome datasets from SARS-CoV infected patient's peripheral blood [44] (GSE1739) and Calu-3 cells [45] (GSE33267) were used. One transcriptome data set of MERS-CoV infected Calu-3 cells [46] (GSE122876) was selected. In addition, one SARS-CoV-2 specific proteomic dataset was collected from a website: https://biochem2.com/index.php/22-ibcii/pqc/130-frontpage-pqc#coronavirus. Adjusted *P* value less than 0.01 was defined as differentially expressed genes/proteins. Differential gene expression in cells treated with various drugs were retrieved from the Connectivity Map (CMap) database[47], and were used as gene profiles for the drugs. We calculated an enrichment score (*ES*) for each CoV signature dataset using a previously described method [48]:

$$ES = \begin{cases} ES_{up} - ES_{down}, & sgn(ES_{up}) \neq sgn(ES_{down}) \\ 0, & else \end{cases} \quad (3)$$

$ES_{up}$ and $ES_{down}$ are calculated separately for the up- and down-regulated genes from the CoV gene signature dataset. We computed $a_{up/down}$ and $b_{up/down}$ as:

$$a = \max_{1 \leq j \leq s} \left( \frac{j}{s} - \frac{V(j)}{r} \right) \quad (4)$$

$$b = \max_{1 \leq j \leq s} \left( \frac{V(j)}{r} - \frac{j-1}{s} \right) \quad (5)$$

where $j = 1, 2, \cdots, s$ are the genes of HCoV signature dataset sorted in ascending order by their rank in the gene profiles of the drug being computed. The rank of gene $j$ is denoted by $V(j)$, where $1 \leq V(j) \leq r$, with $r$ being the number of genes (12,849) from the CMap database. Then, $ES_{up/down}$ is set to $a_{up/down}$ if $a_{up/down} > b_{up/down}$, and is set to $-b_{up/down}$ if $b_{up/down} > a_{up/down}$. To quantify the significance of the *ES* scores, permutation tests are repeated 100 times using randomly generated gene lists with the same number of up- and down-regulated genes as the CoV signature data set. Drugs are considered to have significantly enriched effect if *ES* > 0 and *P* < 0.05.

**Performance Evaluation**

We introduced the area under the receiver operating characteristic (ROC) curve (AUROC) several evaluation metrics for evaluating performance of drug-target interaction prediction. The area under the receiver operating characteristic (ROC) curve (AUROC)[49] is the global prediction performance. The ROC curve is obtained by calculating the true positive rate (TPR) and the false positive rate (FPR) via varying cutoff.

## Supplemental Information

Supplemental Information can be found online.

## Data and Code Availability.

Source code and data can be downloaded from https://github.com/ChengF-Lab/CoV-KGE.

## Acknowledgements

We acknowledge support from the Amazon Cloud, for credits to AWS ML Services.

**Conflict of interest:** The content of this publication does not necessarily reflect the views of the Cleveland Clinic. The authors declare no competing interests.

## REFERENCES

1. Dong, E., Du, H. & Gardner, L. An interactive web-based dashboard to

track COVID-19 in real time. *Lancet Infect Dis. doi: 10.1016/S1473-3099(20)30120-1* (2020).

2. Sanders, J.M., Monogue, M.L., Jodlowski, T.Z. & Cutrell, J.B. Pharmacologic treatments for coronavirus disease 2019 (COVID-19): A Review. *JAMA* doi:10.1001/jama.2020.6019 (2020).

3. Avorn, J. The $2.6 billion pill--methodologic and policy considerations. *N Engl J Med.* **372**, 1877-1879 (2015).

4. Harrison, C. Coronavirus puts drug repurposing on the fast track. *Nat Biotechnol.* **38**, 379-381 (2020).

5. Cao, B. et al. A trial of lopinavir-ritonavir in adults hospitalized with severe Covid-19. *N Engl J Med.* doi: 10.1056/NEJMoa2001282 (2020).

6. Hoffmann, M. et al. SARS-CoV-2 cell entry depends on ACE2 and TMPRSS2 and is blocked by a clinically proven protease inhibitor. *Cell* **181**, 271-280 (2020).

7. Gordon, D.E. et al. A SARS-CoV-2-human protein-protein interaction map reveals drug targets and potential drug-repurposing. *bioRxiv*, https://www.biorxiv.org/content/10.1101/2020.03.22.002386v3 (2020).

8. Cheng, F. et al. Systems biology-based investigation of cellular antiviral drug targets identified by gene-trap insertional mutagenesis. *PLoS Comput Biol* **12**, e1005074 (2016).

9. Zhou, Y. et al. Network-based drug repurposing for novel coronavirus 2019-nCoV/SARS-CoV-2. *Cell Discov* **6**, 14 (2020).

10. Cheng, F. et al. Network-based approach to prediction and population-based validation of in silico drug repurposing. *Nat Commun* **9**, 2691 (2018).

11. Cheng, F., Kovacs, I.A. & Barabasi, A.L. Network-based prediction of drug combinations. *Nat Commun* **10**, 1197 (2019).

12. Cheng, F. et al. A genome-wide positioning systems network algorithm for in silico drug repurposing. *Nat Commun* **10**, 3476 (2019).

13. Wang, M. et al. Deep Graph Library: Towards Efficient and Scalable Deep Learning on Graphs. ICLR Workshop on Representation Learning on Graphs and Manifolds. https://arxiv.org/pdf/1909.01315.pdf (2019).

14. Percha, B. & Altman, R.B.J.B. A global network of biomedical relationships derived from text. Bioinformatics, **34**, 2614-2624 (2018).

15. Wishart, D.S. et al. DrugBank 5.0: a major update to the DrugBank database for 2018. *Nucleic Acids Res* **46**, D1074-D1082 (2018).

16. Sun, Z., Deng, Z., Nie, J. & Tang, J. ROTATE: Knowledge graph embedding by relational rotation in complex space. ICLR, 2019. https://arxiv.org/pdf/1902.10197.pdf (2019).

17. van der Maaten, L. & Hinton, G. Visualizing data using t-SNE. *J Mach Learn Res.* **9**, 2579-2605 (2008).

18. Chen, G. et al. Clinical and immunologic features in severe and moderate Coronavirus Disease 2019. *J Clin Invest.* (2020).

19. Wang, X. et al. SARS-CoV-2 infects T lymphocytes through its spike protein-mediated membrane fusion. *Cell Mol Immunol.* (2020).

20. Amici, C. et al. Indomethacin has a potent antiviral activity against SARS coronavirus. *Antivir Ther.* **11**, 1021-1030 (2006).

21. Xu, T., Gao, X., Wu, Z., Selinger, W. & Zhou, Z. Indomethacin has a potent antiviral activity against SARS CoV-2 in vitro and canine coronavirus in vivo. *bioRxiv* https://doi.org/10.1101/2020.04.01.017624 (2020).

22. Tan, D.X., Korkmaz, A., Reiter, R.J. & Manchester, L.C. Ebola virus disease: potential use of melatonin as a treatment. *J Pineal Res.* **57**, 381-384 (2014).

23. Boga, J.A., Coto-Montes, A., Rosales-Corral, S.A., Tan, D.X. & Reiter, R.J. Beneficial actions of melatonin in the management of viral infections: a new use for this "molecular handyman"? *Rev Med Virol.* **22**, 323-338 (2012).

24. Wang, D. et al. Clinical characteristics of 138 hospitalized patients with 2019 novel coronavirus–infected pneumonia in Wuhan, China. *JAMA* **323**, 1061-1069 (2020).

25. Lasso, G. et al. A Structure-informed atlas of human-virus interactions. *Cell* **178**, 1526-1541.e1516 (2019).

26. Zhao, Y. et al. Toremifene interacts with and destabilizes the Ebola virus glycoprotein. *Nature* **535**, 169-172 (2016).

27. Johansen, L.M. et al. FDA-approved selective estrogen receptor modulators inhibit Ebola virus infection. *Sci Transl Med.* **5**, 190ra179 (2013).

28. Cong, Y. et al. MERS-CoV pathogenesis and antiviral efficacy of licensed drugs in human monocyte-derived antigen-presenting cells. *PLoS One* **13**, e0194868 (2018).

29. Dyall, J. et al. Repurposing of clinically developed drugs for treatment of Middle East respiratory syndrome coronavirus infection. *Antimicrob Agents Chemother.* **58**, 4885-4893 (2014).

30. Jeon, S. et al. Identification of antiviral drug candidates against SARS-CoV-2 from FDA-approved drugs. *bioRxiv. doi: https://doi.org/10.1101/2020.03.20.999730* (2020).

31. Valavaara, R. et al. Toremifene, a new antiestrogenic compound, for treatment of advanced breast cancer. Phase II study. *Eur J Cancer Clin Oncol.* **24**, 785-790 (1988).

32. Thompson, I.M., Jr. & Leach, R. Prostate cancer and prostatic intraepithelial neoplasia: true, true, and unrelated? *J Clin Oncol.* **31**, 515-516 (2013).

33. Kivinen, S. & Maenpaa, J. Effect of toremifene on clinical chemistry, hematology and hormone levels at different doses in healthy postmenopausal volunteers: phase I study. *J Steroid Biochem.* **36**, 217-220 (1990).

34. Liu, J. et al. Hydroxychloroquine, a less toxic derivative of chloroquine, is effective in inhibiting SARS-CoV-2 infection in vitro. *Cell Discov.* **6**, 16 (2020).

35. Chen, Z. et al. Efficacy of hydroxychloroquine in patients with COVID-19: results of a randomized clinical trial. *medRxiv*, https://doi.org/10.1101/2020.03.22.20040758 (2020).

36. Chorin, E. et al. The QT interval in patients with SARS-CoV-2 infection treated with hydroxychloroquine/azithromycin. *medRxiv*, https://doi.org/10.1101/2020.04.02.20047050 (2020).

37. Rajeshkumar, N.V. et al. Fatal toxicity of chloroquine or hydroxychloroquine with metformin in mice. *bioRxiv*, https://doi.org/10.1101/2020.03.31.018556 (2020).

38. Gassen, N.C. et al. SKP2 attenuates autophagy through Beclin1-ubiquitination and its inhibition reduces MERS-Coronavirus infection. *Nat Commun* **10**, 5770 (2019).

39. Shi, J. et al. Susceptibility of ferrets, cats, dogs, and other domesticated animals to SARS-coronavirus 2. *Science* (2020).

40. Chen, G. et al. Clinical and immunologic features in severe and moderate Coronavirus Disease 2019. *J Clin Invest.* (2020).

41. Wishart, D.S. et al. DrugBank 5.0: a major update to the DrugBank database for 2018. **46**, D1074-D1082 (2018).

42. Sun, Z., Deng, Z.-H., Nie, J.-Y. & Tang, J. Rotate: Knowledge graph

embedding by relational rotation in complex space. *ICLR* (2019).

43. Zheng, Da, et al. "DGL-KE: Training Knowledge Graph Embeddings at Scale." arXiv preprint arXiv:2004.08532 (2020).

44. Reghunathan, R. et al. Expression profile of immune response genes in patients with Severe Acute Respiratory Syndrome. *BMC Immunol.* **6**, 2 (2005).

45. Josset, L. et al. Cell host response to infection with novel human coronavirus EMC predicts potential antivirals and important differences with SARS coronavirus. *mBio* **4**, e00165-00113 (2013).

46. Yuan, S. et al. SREBP-dependent lipidomic reprogramming as a broad-spectrum antiviral target. *Nat Commun.* **10**, 120 (2019).

47. Lamb, J. et al. The Connectivity Map: using gene-expression signatures to connect small molecules, genes, and disease. *Science* **313**, 1929-1935 (2006).

48. Sirota, M. et al. Discovery and preclinical validation of drug indications using compendia of public gene expression data. *Sci Transl Med.* **3**, 96ra77 (2011).

49. Powers, D.M. Evaluation: from precision, recall and F-measure to ROC, informedness, markedness and correlation. *J Mach Learn Technol.* **2**, 37-63 (2011) methodology (termed CoV-KGE) for drug repurposing in COVID-19. Specifically, a comprehensive knowledge graph that contains 15 million edges across 39 types of relationships connecting drugs, diseases, genes, pathways, and expressions, and others by incorporating data from 24 million PubMed publications (Supplemental Table S2). Subsequently, a deep learning approach (RotatE in DGL-KE) was used to prioritize high-confidence candidate drugs for COVID-19 under Amazon supercomputing resources (*cf.* Methods section). Finally, all CoV-KGE predicted drug candidates were future validated by three gene expression datasets in SARS-CoV-1 infected human cells and one proteomic data in SARS-CoV-2 infected human cells.

**Figure Legends**

**Figure 1.** A diagram illustrating workflow of a network-based, deep learning

**Figure 2**. The performance of CoV-KGE in predicting of drug canidates for COVID-19. Drugs under the ongoing COVID-19 trial data (https://covid19-trials.com/) were used as the validation set. AUROC: the area under the ROC curve.

**Figure 3**. A diagram illustrating the landscape of CoV-KGE predicted repurposable drugs for COVID-19. (**A**) Visualization of the drug vector learned by knowledge graph embedding using the t-SNE (t-distributed stochastic neighbor embedding algorithm[17]). The two-dimensional (2D) representation of the learned vectors for 14 types of drugs grouped by the first-level of the Anatomical Therapeutic Chemical (ATC) classification system codes.

Semantically similar ATC drugs are mapped to nearby representations. We highlight the 11 drugs which are under clinical trials for COVID-19 and 3 experimentaly drugs (toremifine, niclosamide and indomethasin) having striking *in vitro* antiviral activitities across Ebola virus[26, 27], MRES-CoV[28], SARS-CoV-1[29], and SARS-CoV-2[30]. (**B**) A proposed mechanism-of-action of hydroxychloroquine in potential treatment of COVID-19. Hydroxychloroquine has both anti-inflammatory[10] and antiviral activity[34]. Given the up-regulation of systemic inflammation (e.g., cytokine storms) observed in severe COVID-19 patients[40], combination therapy with antiviral agents (toremifene or niclosamide) and with anti-inflammatoray agents (melatonin or indomethacin) may offer promising treatment approach for COVID-19, which is warranted to be tested in the randomized controlled trials further.

# Figure 1

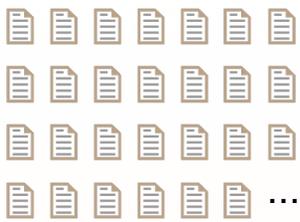
**24 million research articles**

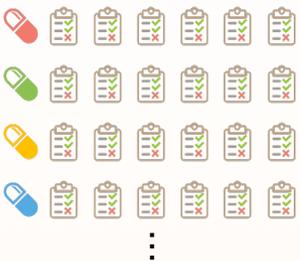
**Knowledge of drugs**

**Drug-Disease**
treatment/therapy (including investigational)
inhibits cell growth
⋮
alleviates, reduces
role in disease pathogenesis
(disease progression) biomarkers

**Gene-Disease**
causal mutations
mutations affecting disease course
drug targets
⋮
biomarkers (diagnostic)
overexpression in disease
improper regulation linked to disease

**Drug-Gene**
binding, ligand (esp. receptors)
increases expression
decreases expression
⋮
metabolism, pharmacokinetics inhibits
antagonism, blocking
agonism, activation

**Gene-Gene** 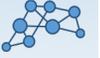
binding, ligand (esp. receptors)
enhances response
activates, stimulates
⋮
same protein or complex
regulation
production by cell population

**RotatE on knowledge graph**

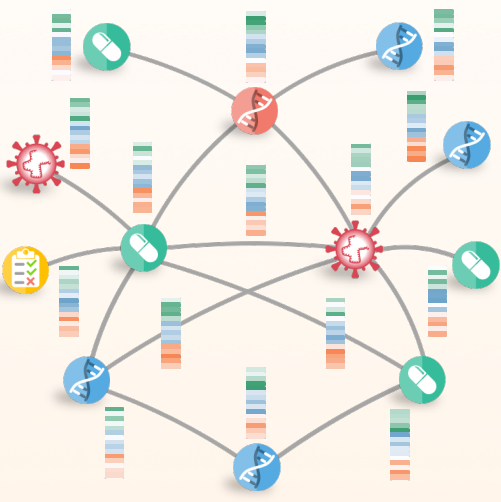

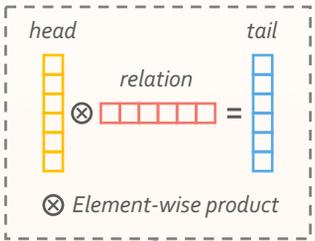
*representation learning*

**39 types of relations and over 15 million edges**

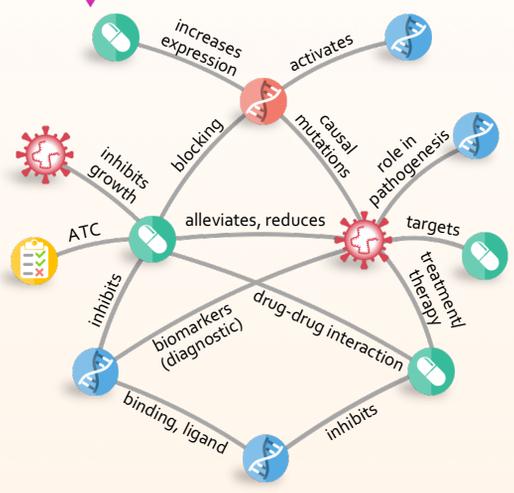

- Drugs (37,112)
- Drug properties (11,289)
- COVID-19
- Genes (89,159)
- HCoV-related genes (120)

$$score = -|h \otimes r - t|^2$$

**top candidates for HCoV-related genes**

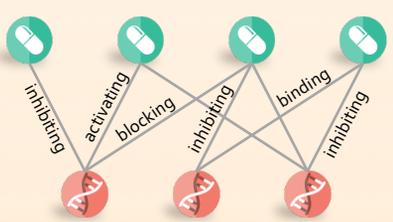

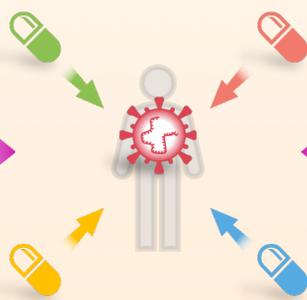

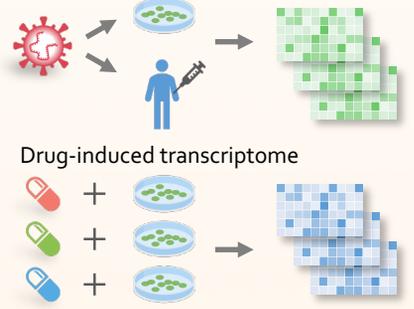

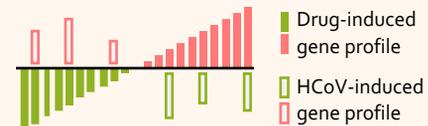

**Enrichment analysis**

**Figure 2**

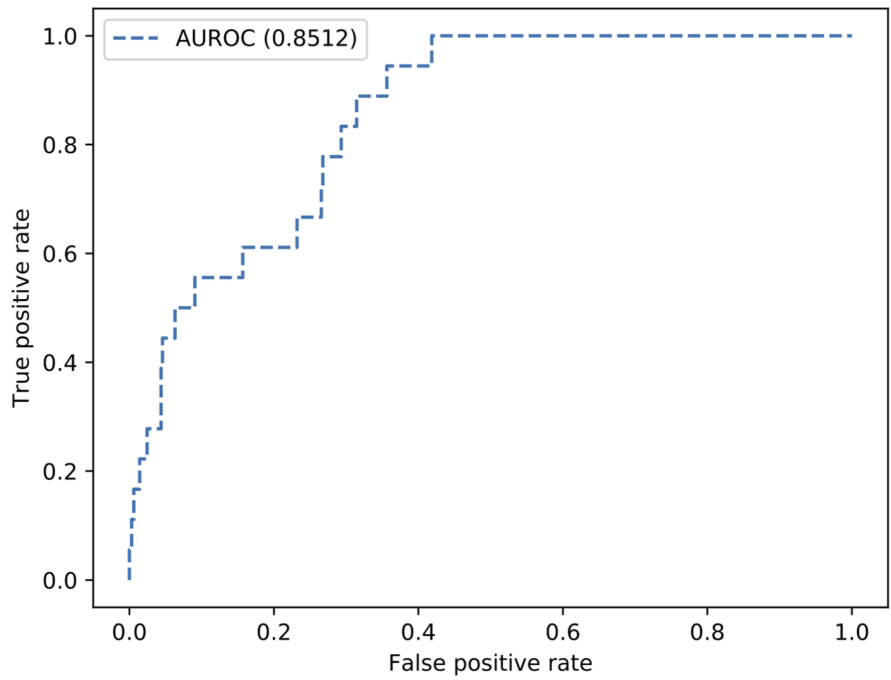

# Figure 3

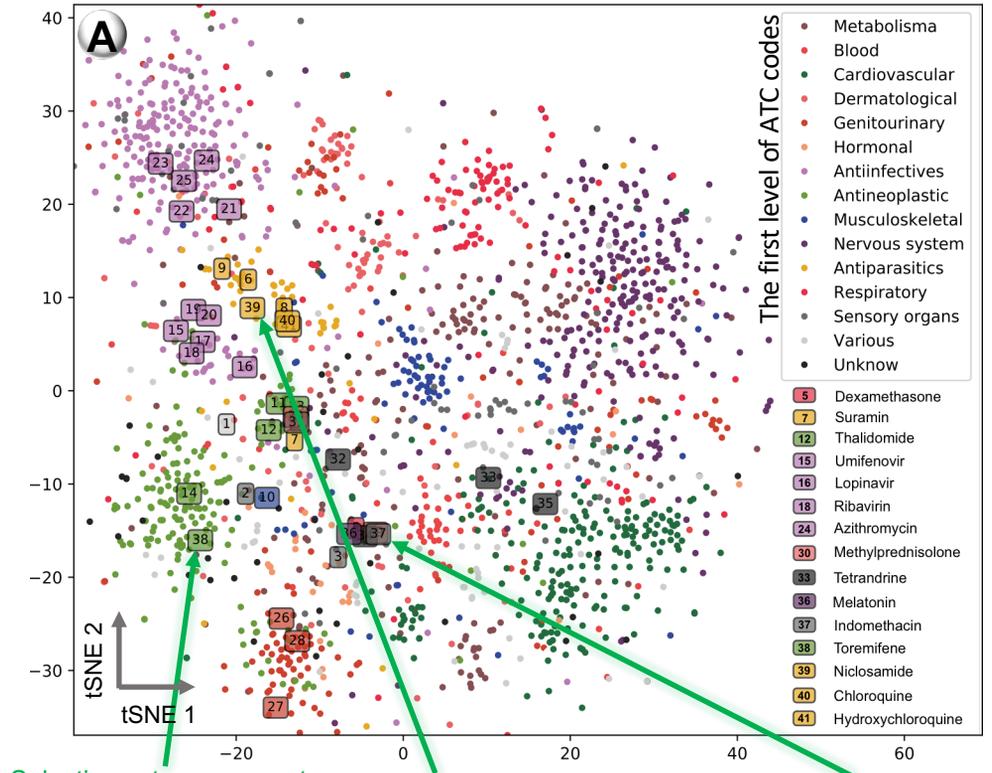
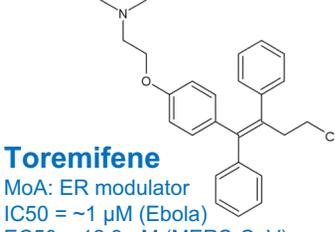
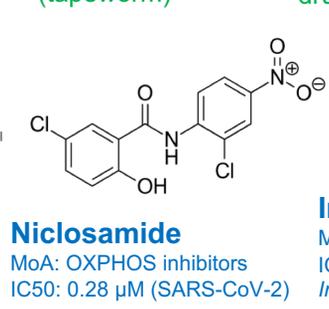
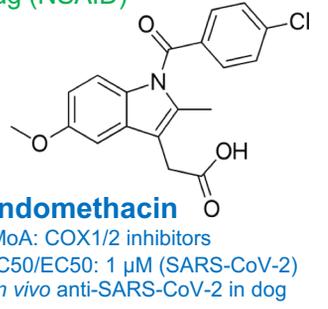
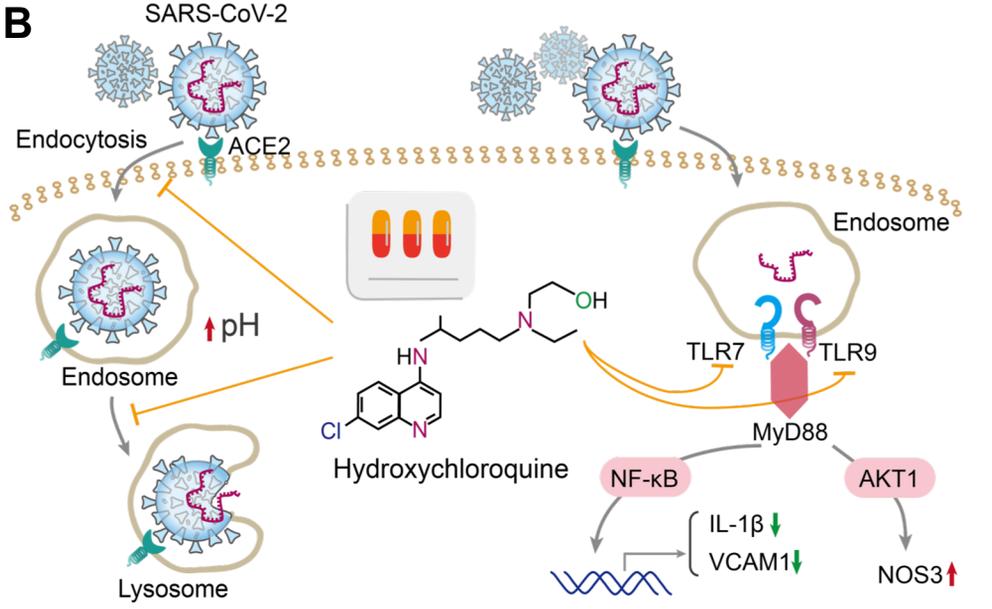